\documentclass[%
 reprint,
groupedaddress,
showpacs,preprintnumbers,
amsmath, amssymb, aps,
]{revtex4-1}

\usepackage{graphicx}% Include figure files
\usepackage{dcolumn}% Align table columns on decimal point
\usepackage{bm}% bold math
\newcommand{\ket}[1]{\ensuremath{\left|#1\right\rangle}}

\begin{document}

%\preprint{APS/123-QED}

\title{Tailoring the correlation and anti-correlation behavior of path-entangled photons in Glauber-Fock oscillator lattices}% Force line 

\author{Armando Perez-Leija$^{1}$, Robert Keil$^{3}$, Alexander Szameit$^{3}$, Ayman F. Abouraddy$^{1}$, Hector Moya-Cessa$^{2}$, and Demetrios N. Christodoulides$^{1}$}
%\altaffiliation[Also at ]{INAOE, Coordinacion de Optica, A.P. 51 y 216, 72000 Puebla, Puebla, Mexico}\email[]{aleija@creol.ucf.edu}
\affiliation{CREOL$/$College of Optics, University of Central Florida, Orlando, Florida,USA\\INAOE, Coordinacion de Optica, A.P. 51 y 216, 72000 Puebla, Puebla, Mexico\\Institute of Applied Physics, Friedrich-Schiller Universitat Jena, Max-Wien-Platz 1, 07743 Jena, Germany}
\date{\today}
%\email[1]{aleija@creol.ucf.edu}
\begin{abstract}
We demonstrate that single-photon as well as biphoton revivals are possible in a new class of dynamic optical systems-the so-called Glauber-Fock oscillator lattices. In these arrays, both Bloch-like oscillations and dynamic delocalization can occur which can be described in closed form. The bunching and antibunching response of path-entangled photons can be pre-engineered in such coupled optical arrangements and the possibility of emulating Fermionic behavior in this family of lattices is also considered. We elucidate these effects via pertinent examples and we discuss the prospect of experimentally observing these quantum interactions.  
\end{abstract}


\maketitle
The prospect of manipulating and engineering quantum states has become an issue of great importance within the framework of quantum information and computation \cite{bennett,kimble}. Along these lines, several physical settings have been envisioned as viable avenues to achieve this goal. Among them, one may mention trapped-ion arrangements and optical lattices as well as spin systems and quantum dots \cite{bose,kouwenhoven}. While the list of such possibilities keeps increasing with time, quantum optics has so far provided a versatile platform where such ideas can be experimentally realized and tested. As previously indicated, in optics, quantum information processing can be achieved entirely linearly, using simple passive components like beam splitters and phase shifters along with standard photodetectors and single-photon sources\cite{milburn}. In this same optical realm, quantum entanglement can arise as a natural byproduct of photon interactions-a clear manifestation of their particle-wave duality. Perhaps, nowhere this process is more apparent than in the so-called Hong-Ou-Mandel two-photon interference effect\cite{mandel}. In this latter configuration, photon entanglement is made possible via quantum interference-afforded after scattering from a beam-splitter. Lately, optical arrays of evanescently coupled waveguides have been suggested as a possible route toward the implementation of multiport systems with moldable quantum dynamics\cite{politi}. The flexibility offered by such compact and often miniaturized optical $N\times N$ configurations is made possible by the exceptional control achievable these days in microfabrication techniques\cite{szameit,keil}. In this regard, Bloch oscillations of NOON and W entangled states as well as quantum random walks have been theoretically considered and observed in such arrays\cite{agarwal,silberberg,peruzzo}. In addition, the evolution of quantum correlations in both periodic and random (Anderson) lattices has also been investigated\cite{bromberg,lahini}. The possibility of classically emulating Jaynes-Cummings systems on such lattices has also recently been proposed\cite{longhi1}. The question naturally arises as to whether such multiport array systems can be utilized as a means to manipulate quantum states of light.\\ 
In this Letter we investigate the propagation dynamics of non-classical light in a new class of dynamic photonic systems-the so-called Glauber-Fock oscillator lattices. We demonstrate that Bloch-like revivals and dynamic delocalization effects can naturally occur in spite of the fact that the structure itself is semi-infinite and not periodic. Interestingly, these interactions can be described in closed form, from where one can analytically deduce the turning points of these quantum oscillations. More importantly, the bunching and antibunching response of path-entangled biphotons can be pre-engineered in such coupled optical arrangements. Emulating Fermionic dynamics in such arrangements are also considered and compared to those expected from bosonic systems in these same arrays. Finally the possibility of experimentally realizing such Glauber-Fock oscillator lattices is discussed.\\
We begin our analysis by considering a semi-infinite Glauber-Fock oscillator array consisting of evanescently coupled waveguides. In this arrangement the coupling coefficients among neighboring channels vary with the square root of the site index, i.e., $C_{k,k+1}\propto \sqrt{n+1}$\cite{keil}. For generality, we also allow this coupling to depend on the propagation distance in this lattice, in which case  $C_{k,k+1}\propto f(Z)\sqrt{n+1}$ where $f(Z)$ is an arbitrary real function. In addition we also assume that the propagation constant (local eigenvalue) of each waveguide element varies linearly with the site position. In essence, in this arrangement the refractive index is linearly increasing-in a way analogous to that of an externally biased crystal in solid state physics.  Starting from these premises, one can show that in this class of arrays, the Heisenberg equation of motion for the creation operator of a single photon in waveguide mode $k$ is given by:

\begin{equation}
\label{eq1}
i\frac{da^{\dag}_{k}}{dZ}-f(Z)\left(\sqrt{k+1}a^{\dag}_{k+1}+\sqrt{k}a^{\dag}_{k-1}\right)-\lambda k a^{\dag}_{k}=0    
\end{equation}

In the above equation, $Z$ represents a normalized propagation distance, and $\lambda$  is a real constant associated with the strength of the aforementioned linear index change among adjacent sites. We emphasize that unlike standard infinite Bloch oscillator arrays\cite{kenkre,lederer,morandoti}, the proposed structure is semi-infinite and asymmetric, e.g. the waveguides are no longer equidistant. As we will see later, these additional degrees of freedom may enable one to observe Bloch-like oscillations even in the neighborhood of the array boundary ($k=0$). This is in contradistinction to the well-known Dunlap-Kenkre system, the only other integrable oscillator lattice\cite{kenkre}. 

In general, the quantum dynamics in this Glauber-Fock oscillator array can be described through the evolution matrix $\overline{T}$  relating the input-output states, i.e., 

\begin{equation}
\label{eq2}
a_{k}^{\dag}(0)=\sum_{n=0}^{\infty}T_{k,n}^{*}(Z)a_{n}^{\dag}(Z).
\end{equation}

\begin{figure}[b!]
\begin{center}
\includegraphics[width=3.5in]{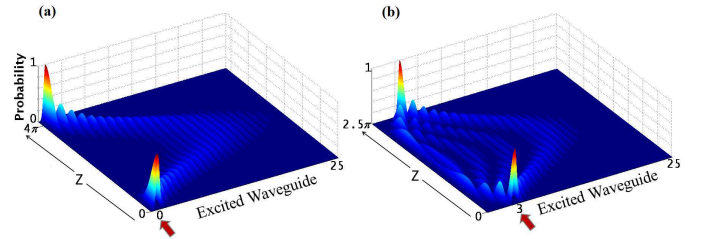}
\caption{(Color online) Bloch-like oscillation for a single photon propagating through Glauber-Fock oscillator lattices. The red arrow indicates the site where the photon is being launched. The used lattice parameters are $\gamma=1$, and $\lambda=1/2$, $\lambda=4/5$ for (a) and (b) respectively.}
\label{f1}
\end{center}
\end{figure}

In Eq.(2), $T_{k,n}^{*}\left(Z\right)$ represents the Hermitian conjugate of the $(k, n)$ element of $\overline{T}$ matrix or unitary transformation. We would like to emphasize that in the present case, the evolution matrix cannot be simply obtained from $\exp(-iZH)$ since the Hamiltonian of the problem is $Z$ (or time) dependent\cite{louisell}. Yet, in spite of this complexity, one can show that the evolution elements $T_{k,n}^{*}$ of the system can be obtained in closed form.  Starting from Eq.(1), one can show that these elements are given by:    

\begin{equation}
\label{eq3}
T_{k,n} =
\left\{
\begin{array}{ll}
\sqrt{\frac{n!}{k!}}e^{(-A-i\lambda nZ)}\left[B\right]^{k-n}L_{n}^{k-n}\left(|B|^{2}\right), &  n \leq k \\
\sqrt{\frac{k!}{k!}}e^{(-A-i\lambda kZ)}\left[C\right]^{n-k}L_{n}^{n-k}\left(|B|^{2}\right), &  n \geq k
\end{array}
\right.
\end{equation}

and $A\left(Z\right)= \int_{0}^{Z}\left(\int_{0}^{Z''} e^{i\lambda(Z'-Z'')}f\left(Z'\right)dZ'\right)f\left(Z''\right)dZ''$, $B\left(Z\right)= -i\int_{0}^{Z} e^{-i\lambda Z'}f\left(Z'\right)dZ'$, and $C\left(Z\right)= -e^{-i\lambda Z}B^{*}\left(Z\right)$.
%\begin{eqnarray}
%&&A\left(Z\right)= -\int_{0}^{Z}\left(\int_{0}^{Z''} %e^{i\lambda(Z'-Z'')}f\left(Z'\right)dZ'\right)f\left(Z''\right)dZ'',\nonumber\\
%&&B\left(Z\right)= -i\int_{0}^{Z} e^{-i\lambda Z'}f\left(Z'\right)dZ'\\
%&&C\left(Z\right)= -e^{-i\lambda Z}B^{*}\left(Z\right).\nonumber
%\end{eqnarray} 
In Eqs.\eqref{eq3} $L_{n}^{m}\left(x\right)$ represents the associated Laguerre polynomials. In addition, Eqs.\eqref{eq3} imply that $\sum_{n=0}^{\infty}\left|T_{k,n}\right|^{2}=1$, in agreement with the fact that  $\overline{T}$ is itself a unitary transformation. 
In order to gain insight into the quantum dynamics in this class of arrays, let us first consider the case where only a single photon is launched into the $k$th waveguide element. We begin by analyzing here the simplest possible scenario where the Hamiltonian of the system is  $Z-$independent, thus $f\left(Z\right)=\gamma$, where  $\gamma$ is a real constant.  Under these conditions, the evolution matrix elements are given by

\begin{equation}
\label{eq5}
T_{k,n} =
\left\{
\begin{array}{ll}
\sqrt{\frac{n!}{k!}}e^{(\delta-i\lambda nZ)}\left[\Theta \right]^{k-n}L_{n}^{k-n}\left(\Phi\right), &  n \leq k \\
\sqrt{\frac{k!}{n!}}e^{(\delta-i\lambda kZ)}\left[\Theta\right]^{n-k}L_{n}^{n-k}\left(\Phi\right), &  n \geq k
\end{array}
\right.
\end{equation}

where $\delta = i\gamma^{2}\lambda^{-1}+\gamma^{2}\left[\exp(-i\lambda Z)-1\right]\lambda^{-2}$, $\Theta=\left(\gamma/\lambda\right)\left[\exp(-i\lambda Z)-1\right]$, and $\Phi = \left(2\gamma^{2}/\lambda^{2}\right)\left[1-\cos\left(\lambda Z\right)\right]$. In this case the probability of finding this single-photon at waveguide site $n$ when launched at $k$, can be obtained from $P_{n,k}\left(Z\right)=\langle a_{n}^{\dag}a_{n}\rangle=\left|T_{k,n}\right|^{2}$. Equations \eqref{eq5} clearly indicate that the associated probability distribution exhibits revivals at regular intervals, e.g. at $Z=2\pi s/\lambda$ ($s$ being integer). At these revival points, for single-photon excitation, all the $T_{k,n}$ coefficients of Eq.\eqref{eq5} vanish except for $n=k$ , i.e., the probability collapses  into the initial waveguide site $k$. Figure \ref{f1} depicts this process for such a Glauber-Fock oscillator array for two different values of $\lambda$. These side views clearly show the previously mentioned "collapse" and revivals of the probability at $Z=2\pi s/\lambda$ irrespective of the site where the photon was initially coupled. In all cases, these Bloch-like oscillations occur and the photon does not escape into the bulk region-towards the right. This in spite of the fact that the optical potential linearly increases and the waveguide elements get physically closer towards higher values of $n$. We note that unlike standard Bloch oscillations occurring in periodic lattices \cite{kenkre,lederer,morandoti}, in this system the dynamics are asymmetric. This broken symmetry is a result of the semi-infinite nature of this particular array. 

\begin{figure}[t!]
\begin{center}
\includegraphics[width=3.5in]{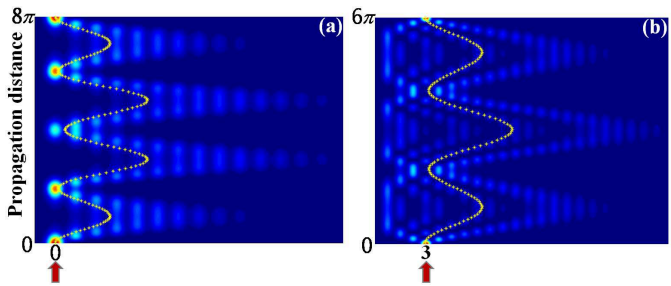}
\caption{(Color online) Evolution of the probability distributions for single photons propagating through this Glauber-Fock oscillator lattices with $\lambda=1$  and  $\varpi=3/4$ for (a), and $\varpi=2/3$ for (b). Dashed lines show the evolution of $|B(Z)|^{2}$ along $Z$, predicting the corresponding revival distance at $8\pi$, and $6\pi$, respectively.}
\label{f2}
\end{center}
\end{figure}

\begin{figure}[t!]
\begin{center}
\includegraphics[width=3.3in]{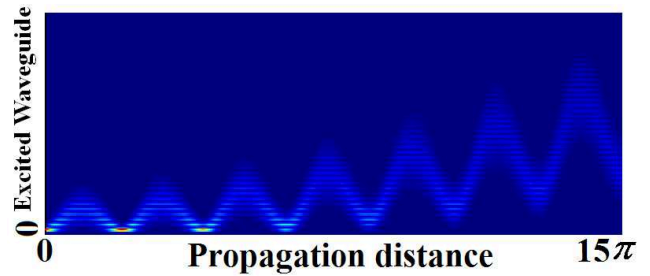}
\caption{(Color online) Theoretical evolution of the probability of having one photon at waveguide $n$ when launched at channel $0$, $P_{n,0}(Z)=\langle a_{n}^{\dag}a_{n}\rangle$, under the resonance condition.}
\label{f3}
\end{center}
\end{figure}

We now focus our attention on the case where the coupling coefficients are  $Z$-dependent, that is when the Glauber-Fock oscillator is dynamic. For illustration purposes we consider the periodic variation: $f(Z)=\kappa_{0}+\epsilon \cos(\varpi Z)$,  where $\kappa_{0}$  is a constant, $\epsilon$ is a coupling modulation amplitude, and $\varpi$ stands for the modulation frequency along the propagation direction.  In this dynamic environment, when a single photon is launched into the $k$-site, the probability will periodically "collapse" into the initial waveguide, at exactly the first zero $\widehat{Z}$ of the function $|B(Z)|^{2}$. For the particular example examined here these revivals occur when the ratio $\varpi/\lambda = P/Q$ is a rational number, and $P$, $Q$  are relatively prime integers. This condition is necessary for the two oscillatory processes independently occurring in this array to lock together synchronously. From here one can deduce that  $\widehat{Z}=2P\pi/\varpi$. This behavior is explicitly illustrated in Fig.\ref{f2} for  $\kappa_{0}=1$, $\lambda=1$, and $\epsilon=0.2$ for the cases $\varpi=3/4$, $2/3$, in (a), adn (b) respectively. The dashed lines on the other hand represent the evolution of the $|B(Z)|^{2}$ function which dictates the period of oscillations. Note that exact revivals do not occur if the ratio $\varpi/\lambda$ is irrational. On the other hand at resonance $\varpi=\lambda$, \emph{dynamic delocalization} occurs. In this regime $|B(Z)|^{2}\propto Z$, and hence a drift motion is induced towards higher site indices \cite{ramy}. Therefore no pure oscillatory behavior is possible at resonance. This delocalization process at resonance is depicted in Fig. \ref{f3}, where it is shown that the probability of finding the photon among the waveguides gradually shifts towards the upper side of array oscillator. 

We next consider the quantum dynamics of entangled pair of photons, launched either spatially correlated or anti-correlated into this class of Glauber-Fock oscillator arrays. As we will see, this new class of photonic lattices can tailor the bunching and antibunching behavior of path-entangled biphotons (spatially extended state). Conceptually, photon pairs (biphotons) with correlated positions will couple into the same waveguide (within a certain excitation window $W$) with an equal probability. In this regime, the corresponding input state is written as $\ket{\psi_{C}}=\sqrt{1/W}\left[\left(a_{f}^{\dag}\right)^{2}+\left(a_{f+1}^{\dag}\right)^{2}+...+\left(a_{l}^{\dag}\right)^{2}\right]\ket{0}$. Throughout our paper, $f$, and $l$ will represent the first and last waveguide site within the excitation window. This input state can be generated by placing the waveguide array immediately after a type I collinear degenerate narrow-band spontaneous parametric down conversion (SPDC) thin-crystal source\cite{byer}. On the other hand, an entangled pair of photons with anti-correlated positions corresponds to the physical situation where the photon pair is always coupled to waveguides on opposite sides of the excitation window $W$-again with equal probability. Thus, the input state is written as $\ket{\psi_{A}}=\sqrt{2/W}\left[a_{f}^{\dag}a_{l}^{\dag}+a_{f+1}^{\dag}a_{l-1}^{\dag}+...+a_{R}^{\dag}a_{R'}^{\dag}\right]\ket{0}$, where $W$ is even and $(R,R')$ represent the floor and ceiling integers of the quantity $(l+f)/2$. In order to obtain the correlation between the array modes, we analyze at the output the coincidence rate at waveguides $p$ and $q$, which is given by $\Gamma_{p,q}\equiv\langle a_{p}^{\dag}a_{q}^{\dag}a_{q}a_{p}\rangle$. In this case one can show that for correlated inputs $\ket{\psi_{C}}$, the correlation map is described by $\Gamma_{p,q}=\left|\sum_{k=f}^{l}T_{p,k}T_{q,k}\right|^{2}$, whereas for anti-correlated $\ket{\psi_{A}}$ it becomes $\Gamma_{p,q}=\frac{1}{2}\left|\sum_{k=0}^{W-1}T_{p,f+k}T_{q,l-k}\right|^{2}$. In order to demonstrate these effects and for comparison purposes, we will always assume here that $W=10$ with the excitation contained between $(f,l)=(0,9)$.   
 
\begin{figure}[t!]
\begin{center}
\includegraphics[width=2.5in]{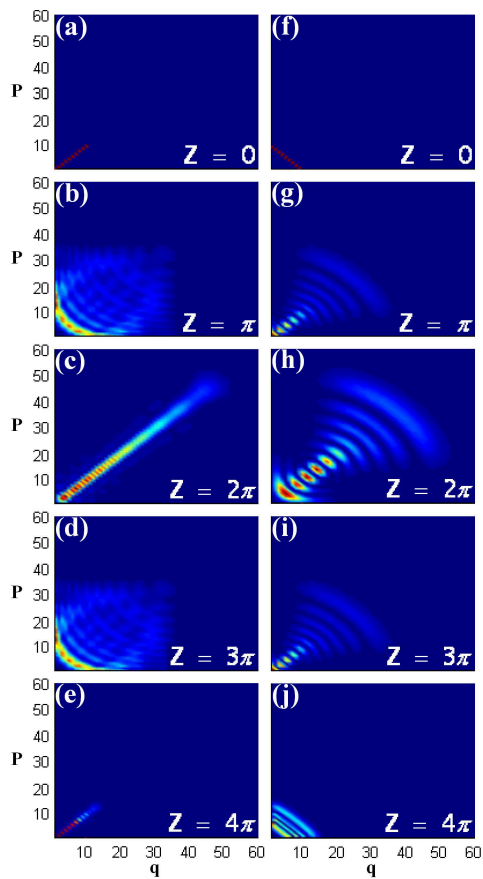}
\caption{(Color online) Left column shows the quantum correlation maps when the input state $\ket{\psi_{C}}$  is coupled into a modulated-tilted Glauber-Fock oscillator array $(\lambda=1/2,\varpi=3/4)$. Similarly, right column  depicts the correlation evolution for the input state$\ket{\psi_{A}}$.}
\label{f4}
\end{center}
\end{figure}

The evolution of the correlation map $\Gamma_{p,q}$ as a function of distance when $f(Z)=1$ and $\lambda=0.5$ is then examined. In this case, revivals are expected at multiples of $Z=4\pi$. When the input state is initially correlated, the map flips and antibunching occurs at $Z=\pi$, and $Z=3\pi$ while it returns to a broadened bunched state in the middle of a cycle. This situation is altered when an anti-correlated biphoton input is used. In this case the map tends to flip over to that of a bunched state at $Z=\pi$, and $Z=3\pi$ while in the middle of this oscillation it attains a correlation mixture-with bunching being predominant. This evolution is altogether different from that occurring in uniform lattices\cite{bromberg}. The present dynamics is a direct outcome of the revivals and of the phase acquired upon reflection from the boundary of this semi-infinite Glauber-Fock oscillator array-which is absent in periodic arrays. We next consider the evolution of correlations when two periods are simultaneously involved in the lattice, e.g. when the function $f(Z)$ is periodic. For this example we again take $f(Z)=\kappa_{0}+\epsilon\cos(\varpi Z)$, with $\kappa_{0}=1$, $\lambda=0.5$, $\epsilon=0.2$, and $\varpi=3/4$ in which case the revival period is $8\pi$. For a correlated input $\ket{\psi_{C}}$ the correlation map exhibits periodic transitions from bunching to antibunching Figs.\ref{f4}(a-e). However, at the half-cycle point the bunching is now not entirely complete, Fig.\ref{f4}(e), due to the incomplete revival of the single photon trajectories. This scenario becomes very different when the initial biphoton state $\ket{\psi_{A}}$ is anti-correlated Fig.\ref{f4}(f-j). The correlation dynamics corresponding to both cases are shown up to half a cycle $(4\pi)$. Evidently, right after the origin, bunching is seen to occur (Fig.\ref{f4}(g)) while midway in the cycle signatures of antibunching behavior appear. This latter pattern is different from that obtained before $\left(f(Z)=1\right)$ when only one oscillation frequency was involved in the Glauber-Fock oscillator. 

\begin{figure}[h!]
\begin{center}
\includegraphics[width=3in]{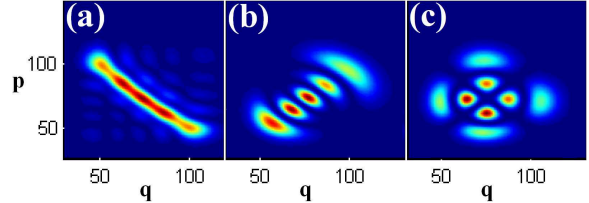}
\caption{(Color online) Correlation matrices at $Z=15\pi$ for the initial states (a) $\ket{\psi_{C}}$, (b) $\ket{\psi_{A}}$ and (c) $\ket{\psi_{F}}$.}
\label{f5}
\end{center}
\end{figure}

We have also explored the response of this system under dynamic delocalization conditions. 
Fig.\ref{f5} depicts again the correlations for the same parameters used in the previous case, except that here $\lambda=\varpi=1$. In this delocalization regime, a correlated input $\ket{\psi_{C}}$  tends to initially oscillate between bunching and antibunching as it was shown in Figs. \ref{f4}(a-c), and eventually settles into antibunched state, Fig.\ref{f5}(a). On the other hand, for an anti-correlated bi-photon input $\ket{\psi_{A}}$ the entangled photons very quickly and irreversibly become bunched and they remain in this state, Fig.\ref{f5}(b). 
The reason why delocalization itself affects the correlation evolution has to do with the fact that in this case the photons tend to eventually escape into the bulk of lattice-away from the boundary. Simulations indicate that by adjusting the two oscillation frequencies one can at will lock the output into a particular bunching/antibunching state. 
In essence the presence of revivals of quantum states (or absence of revivals) allows one to engineer the quantum dynamics in this class of dynamic Glauber-Fock oscillator lattices.\\

Finally, we also consider this same arrangement when Fermionic-like input states are used \cite{matthews}. Figure \ref{f5}(c) shows how the correlation evolves in this case under delocalization conditions for parameters identical to those used in Fig.\ref{f3}. The input in this case is assumed to be of the type $\ket{\psi_{F}}=b_{f}^{\dag}b_{l}^{\dag}\ket{0}$. For this input, the antibunching behavior in the correlation matrix is evident.

In conclusion we have shown that a new family of dynamic arrays, the so-called Glauber-Fock oscillator lattices can be used as a way to mold the quantum evolution of path-entangled photons. In these systems revivals and dynamic delocalization are possible-each leaving a specific mark on the correlation map. If the two oscillation periods associated with these Bloch-like oscillators are irrational with respect to each other, the dynamics become aperiodic. At this point several intriguing questions remain. For example, of interest will be to examine how such structures respond to other maximally entangled states (like NOON states) or whether they can be used to synthesize other quantum states of interest. The response of these lattices may be also useful in considering other classes of problems in other physical configurations having similar quantum analogues like those of the Bose-Hubbard or Jaynes-Cummings type with time varying couplings\cite{longhi1,longhi}. 
\bibliography{GF}
\end{document}